# Performance analysis of Machine learning algorithms for predicting malware

ABM.Adnan Azmee
Computer Science and Engineering
BRAC University
Dhaka, Bangladesh
abm.adnan.azmee@g.bracu.ac.bd

Pranto Protim Choudhury
Computer Science and Engineering
BRAC University
Dhaka, Bangladesh
pranto.protim.choudhury@g.bracu.ac.bd

Md. Aosaful Alam
Computer Science and Engineering
BRAC University
Dhaka, Bangladesh
md.aosaful.alam@g.bracu.ac.bd

Orko Dutta
Computer Science and Engineering
BRAC University
Dhaka, Bangladesh
orko.dutta@g.bracu.ac.bd

Muhammad Iqbal Hossain
Computer Science and Engineering
BRAC University
Dhaka, Bangladesh
iqbal.hossain@bracu.ac.bd

***Abstract*—Malware poses a persistent and evolving threat to modern computing systems, making accurate and timely detection a critical cybersecurity challenge. Traditional signature-based antivirus solutions often fail to identify newly emerging malware, leaving systems vulnerable until updated signatures become available. To address this limitation, this study proposes a machine learning-based malware detection framework capable of distinguishing malicious software from benign applications with high accuracy. Several state-of-the-art classification algorithms, including Artificial Neural Networks (ANN), Support Vector Machines (SVM), XGBoost, and Extra Trees Classifier, were evaluated and compared using a benchmark malware dataset. Experimental results demonstrate that XGBoost achieved the best performance, attaining an accuracy of 98.62%, outperforming the other evaluated models. To demonstrate the practical applicability of the proposed approach, a real-time client–server malware detection system was also developed using the Flask framework, enabling efficient classification of executable files as malicious or benign. The findings highlight the effectiveness of advanced machine learning techniques for enhancing malware detection and contribute toward the development of intelligent and scalable cybersecurity solutions.**

## I. INTRODUCTION

Malware or malicious software, a dangerous computer code, intended to disturb, cripple or take control of a computer system without the approval of the user. The term is utilized to portray a lot of dangers on the web, and it comes in numerous structures, generally covered up in another record or camouflaged as a harmless application. It takes advantage of the technical faults or vulnerabilities in the operating system, hardware, and software. Malware is frequently used to take assets from the PC or attempt to steal significant data, records or cash from an individual.

Most of the time, certain individuals make malware for their own interests and advantages and then it spreads quickly through the internet. Malware has been on the internet for quite a long amount of time and "Brain" is considered to be the first virus in the history on personal computers (pc). It was originated by certain Pakistani young boys in their late teens for their motivations and purposes. The boot area of storage media formatted with the DOS File Allocation Table (FAT) record framework gets infected by the virus. Amjad Farooq Alvi was the mind behind the coding section of Brain. Brain influences the IBM PC by supplanting the boot sector of a floppy disk with a duplicate of the infection. Elk Cloner, another famous malware founded in 1982 and often referred to as the first computer virus, was found on MAC. Elk Cloner was a boot sector virus that spread via floppy discs. The infection was connected to a game and when played the 50th time, the infection would activate, making the screen go dark and showing a poem on the screen.

Not all malware is disastrous, but it can cause certain levels of distress, like it can cause the laptop or computer to slow down or run at a slow pace and can also cause irritating conduct like creating a series of pop-up advertisements. Most of the time, it redirects to different websites and starts playing unwanted videos or songs.

Much research has been conducted on developing new techniques and strategies to assemble, study, and ease noxious code as malware is spreading at an alarming rate on the web. Beyond a shadow of a doubt, it is basic to collect and think about malware found on the Internet. Be that as it may, it is much more essential to create moderation and location strategies in light of the bits of knowledge picked up from the investigation work. Unfortunately, current host-based detection methods undergo from ineffectual detection models. These models focus on the highlights of a particular malware occasion and are regularly effectively evadable by obfuscation or polymorphism. Additionally, finders that check for the nearness of a grouping of framework calls displayed by a malware example are frequently evadable by framework call reordering. To address the inadequacies of weak models, a few powerful discovery approaches have been suggested that expect to recognize the conduct shown by a malware family.

Malware is a program designed solely to cause problems and as days pass more malware is created in addition to the existing old ones. Antivirus is often unaware of any new virus or malware that is being spread through the internet and by the time a solution comes, users have already been affected by the new virus. And this can lead to the loss of useful information and money. Saving personal data is everyone's concern, but in this era of technology, it seems quite impossible. This malicious malware is the reason for losing billions of dollars as well as data. Cybercriminals will take an expected thirty-three billion records by 2023 as indicated by a recent report from Juniper Exploration [5]. Every year billions of data is being stolen from various sites and some of them are used even for bad purposes. Almost sixty million Americans have been influenced by data fraud as per a 2018 online overview by The Harris Survey. A similar overview demonstrates that about fifteen million purchasers experienced data fraud in 2017 [6].

The number of cyber-attacks is increasing daily with an exponential rate. According to the Symantec Internet Security Threat report 2019, the use of destructive malware by different groups increased by about twenty-five in 2018 [7]. A cybersecurity report says that from December 2018 to January 2019 the number of malware activities increased by sixty-one percent [8]. Four thousand ransomware attacks occur daily according to the FBI report [9]. The attacks are causing panic among users, according to a Gallup study, more than seventy percent of Americans are worried about losing personal or financial data by getting hacked [10]. These attacks are also causing great financial loss. In the Accenture report, they say that a company on average costs 2.4 million US dollars due to malware attacks [11].

Many antivirus companies claim that their database is being updated periodically, but it cannot prevent a “zero-day” attack. The signature-based antivirus fails to detect malware in such cases. The effectiveness of antivirus is decreasing day by day because attackers are finding new ways to evade the antivirus. When the antivirus company finds out about a new malware, they need to reverse engineer it and extract the signature which is then added to their database, but by the time it’s added to the database, a lot of users are already under attack. Antivirus vendors are struggling to keep up with attacks. As we see that traditional signature-based antiviruses are not efficient, we propose a malware prediction system based on machine learning. Our system uses the ability of machine learning algorithms to detect and predict different types of malware.

## II. LITERATURE REVIEW

In paper [17], they worked on portable executables and tried to figure out which of the files are malware and created an integrated feature set based on raw and derived inputs. Moreover, they used only 6 algorithms to compare the integrated and raw feature set. Furthermore, they used the 10fold cross-validation technique. They created two datasets with the help of virus-share, Windows XP and Windows 7. One of them had two thousand seven hundred twenty-two malware and two thousand four hundred eighty-eight benign data and another one had one hundred twenty-nine malware data and thirty benign data. Lastly, they changed feature numbers several times and always found that the integrated feature set always gave better results than the raw feature set. In our paper, we use more algorithms than them, and our dataset is more ethical because we collected it from kaggle. And, we have more data than their dataset.

In paper [18], they made a model that was intended to help in the network security of enterprises. They did it because they felt that the signature-based antivirus system could be enough for household purposes but might be a threatening issue for enterprise networks. However, they used only three algorithms to compare with their model and worked on only five thousand data. Their model gave ninety-eight percent accuracy but only worked well for enterprise networks, not with personal computers of home. Also, they did not mention any processing techniques like PCA, LDA but we used some of those techniques in our work. We are also ahead of them in comparison to the highest accuracy rate of algorithms and the amount of data in the dataset.

A PE malware detection system is created by the authors of the paper [19]. Their system worked in real-time. Actually, they worked in three steps, these are feature extraction, selection, and classification. For feature extraction,pefile, which is a python module had been used. But they only used five hundred fifty-two data which is way less than us. Moreover, they used the chi-square test, which found the relation between features with statistics and eventually works well with the small amount of data. If there are lots of data, it can lead to erroneous conclusions whereas our model did well with a large amount of data. We use a variety of algorithms like decision trees, boosting, neural networking. Lastly, their accuracy rate was 97.25%, we found more accuracy in our work even with a large amount of data.
Another real-time malware detection work was done on paper[20]. They extracted only 35 features, again less than our number of features. They used only four algorithms; cnn,mlp,svm and random forest and netmate technique for feature extraction. It is another paper using fewer algorithms and got less accuracy rate than us. Each of their algorithms gave an accuracy rate of more than eighty-five percent. Whereas in our work, we got a more than eighty-five percent accuracy rate in several algorithms. And they did not use any feature reduction technique.
In paper [21], research had been done on machine learning malware detection. In the early stage of our work, we got different ideas from this paper. According to them, the heuristic machine learning is better than signature method which can not stop new malware and dynamic method which has time and space complexity They made the comparison on several steps which are necessary for machine learning malware detection. Firstly, they compare among different feature extraction techniques like signature-based, dll function call, binary sequence, assembly sequence, pe file header, machine activity matrix, entropy signal. Secondly, they showed their survey on feature selection methods like information gain, redundant feature removal, principal component analysis, random forest, self-organizing feature map, wavelet transform. Lastly, they discussed three algorithms, these are supported machine vector, random forest, and artificial neural network and they compare accuracy rate based on feature selection and algorithms.

A hybrid machine learning technique is used for the purpose of malware detection in [22]. They had the target of tracing and categorizing malware and had a better accuracy rate. For gaining that, they collected 5.05 GB of benign along with 1.28 GB malware data and extract of n-gram and pe type of data. They only used three supervised classifiers (J48, Random Forest, Naive Bayes) and one unsupervised classifier (Self-organizing feature map). They used the information gain technique for feature selection and found that random forest gave the highest accuracy rate of ninety-four percent (less than us).

And in paper [23], they created a big dataset and opened it for all research purposes. Their dataset contained three hundred thousand malicious data, three hundred thousand benign data, and three hundred thousand data that were not labeled. For test purposes, they reduced it to one hundred thousand malicious data and three hundred thousand benign data and made another dataset. However, they used eight types of features, only using LightGBM and proved that they gave better performance than another deep learning model, malcov.

## III. Work Plan and Our Proposed Approach

Various approaches used by signature-based antivirus software are not that effective for detecting malware that is not present in their databases. In this section, our proposed model for detecting malware is discussed. First, we need to perform some data preprocessing, which is the process of transforming the raw data into the required format. In the preprocessing part, a label encoding technique is used. Label encoding can be imported from the Sklearn library. It's a significant technique for supervised learning datasets, which encodes categorical data into numeric data. In our model, we used it to transform the label of our target column. We try to check for missing values in the dataset and then split the data set into training and a test set for further processing. Afterward, we perform StandardScaler, which is another data preprocessing technique. The purpose of Standard-Scaler is Standardization; it's going to transform the data into the data so that the standard deviation of the data will be 1 and the mean of the data will be 0. For malware detection, we used Naïve bayes, Logistic regression and finally an artificial neural network (ANN).

For malware detection, at first, Naive Bayes algorithm is used. It is a supervised machine-learning algorithm that is used to classify data into predefined classes. It utilizes the idea of conditional probability to classify by arranging the test information dependent on applying Bayes' theorem as follows: P(A|B) = $P(B|A)\ P(A)\ /P(B)$. It assumes that the effect of a certain feature of a sample is independent of the other features. That implies that each character of a sample contributes freely to decide the probability of the classification of that sample, outputting the category of the highest probability of the sample. Naive Bayes is a classification algorithm appropriate for binary and multiclass classification. Naive Bayes classifiers are mostly used in content arrangement, Spam separating, Sentiment Analysis. In our proposed model we utilized Gaussian Naive Bayes as it is the most effortless to work with on the grounds that we just need to evaluate the mean and the standard deviation from our training data.

Moreover, a logistic regression algorithm is used to predict discrete or categorical values. It is mainly a classification algorithm which is used in cases like fraud detection, email spam detection, etc. where we have to make decisions between yes and no. However, the Logistic curve is not a straight curve like linear regression. It is known as sigmoid curve and here probability p =1/ 1-e^-z where z= mx+c. This equation of probability ensures that predicator will be between 0 and 1. In this algorithm, a threshold value is required where values greater than the threshold are treated as probability 1 and smaller values of threshold are treated as probability 0. In our model after preprocessing by standard scaler, we used logistic regression classifier.

Further, the Artificial Neural Network Algorithm (ANN) is applied, which is developed from the idea of how the neurons of human brains are connected, and how the nervous system performs its work. Artificial Neural Network has many uses and classification is one of them. It can perform well for large datasets; instead of taking the entire dataset it takes data samples. Basically, the artificial neural network has three different layers, the first one is input layer, where data are given as input, the second one is hidden layer, and the final one is the output layer. Depending on work the number of hidden layers varies. In our model we used two hidden layers. We used Rectified Linear Unit (ReLU) and Sigmoid function as activation function in our model. After the model is defined, we compiled it, we used cross-entropy as the loss function. For optimization, we used Adam optimization algorithm that is an extension of greatly used optimization algorithm, stochastic gradient descent.

## IV. Evaluation

In the dataset, there are a total of 100000 samples, among them 50,000 are benign, and 50,000 are malignant. In order to visualize the malware dataset, a heatmap tool is used, which is a two-dimensional graphical representation of the matrix and shows the correlation among the input data. Seaborn visualization library is used to show the heatmap of our input data as shown in Figure 1.

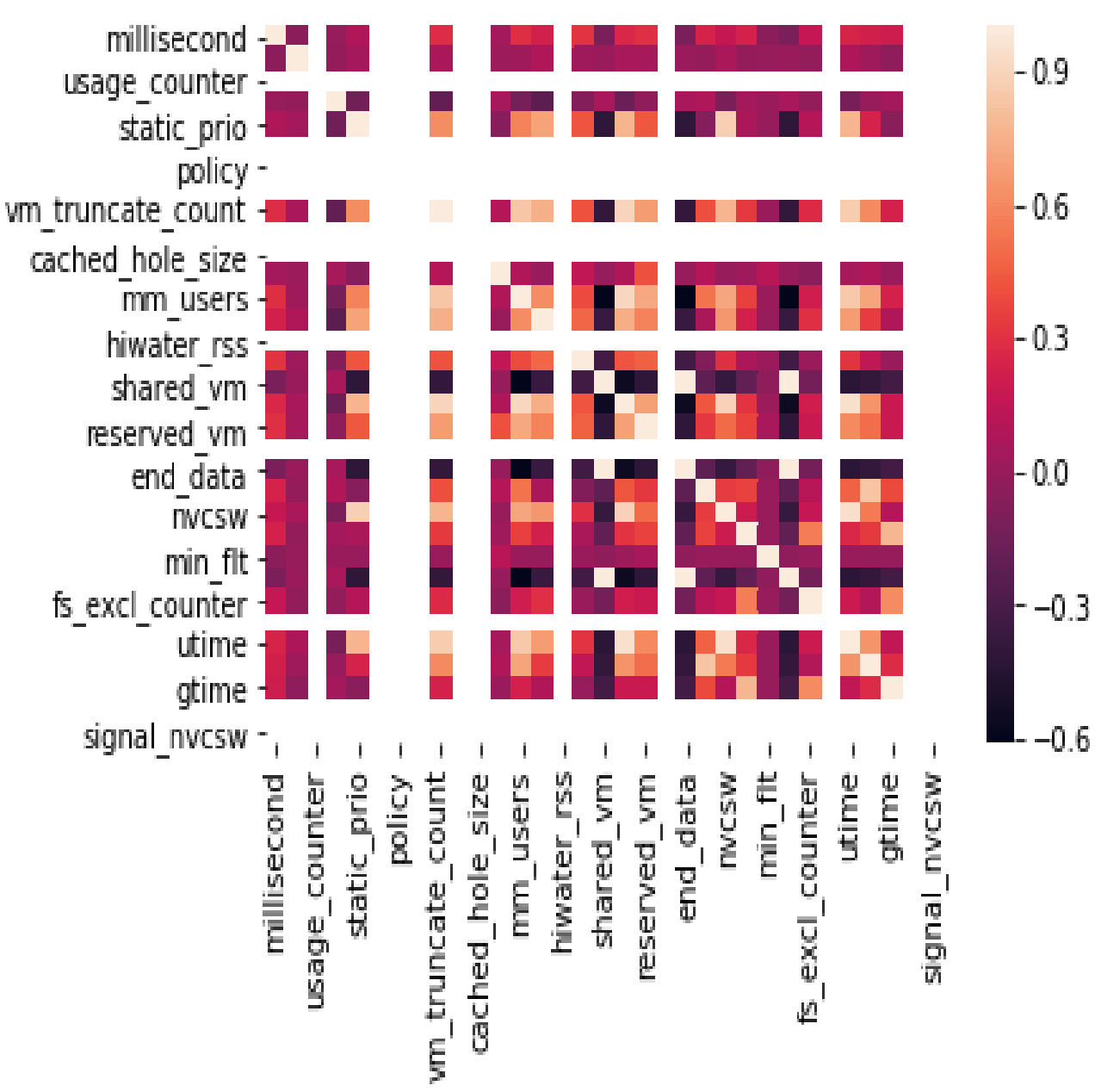

Figure 1. Heatmap of the input data

For evaluating our models, true positive rate (TPR), false-positive rate (FPR), accuracy score, etc metrics are used.

(i) True positive rate: The true positive rate (TPR) is also known as recall it shows the ability of our classifier to find the malicious sample.

TPR= True Positive / (True Positive + False Negative).

(ii) False-positive rate: The false-positive rate (FPR) shows that the possibility of benign sample wrongly classified as a malware sample.

FPR= False Positive / (True Negative + False Positive).

The result of the experiment is shown in table 1. It depicts that, among the three algorithms the artificial neural network algorithm is performing the best.

**Table 1. Experimental results**

| Algorithm | TPR | FPR | Accuracy |
|---|---|---|---|
| Naïve Bayes | 0.93 | 0.53 | 69.5 % |
| Logistic Regression | 0.95 | 0.08 | 93.7 % |
| Artificial Neural Network | 0.99 | 0.021 | 98.6 % |

The receiver operating characteristic (ROC) curve is generated in figure 2 for the three algorithms (RED: Naïve Bayes, GREEN: Logistic regression, YELLOW: ANN) where y-axis plots the true positive rate and the false positive rate at x-axis.

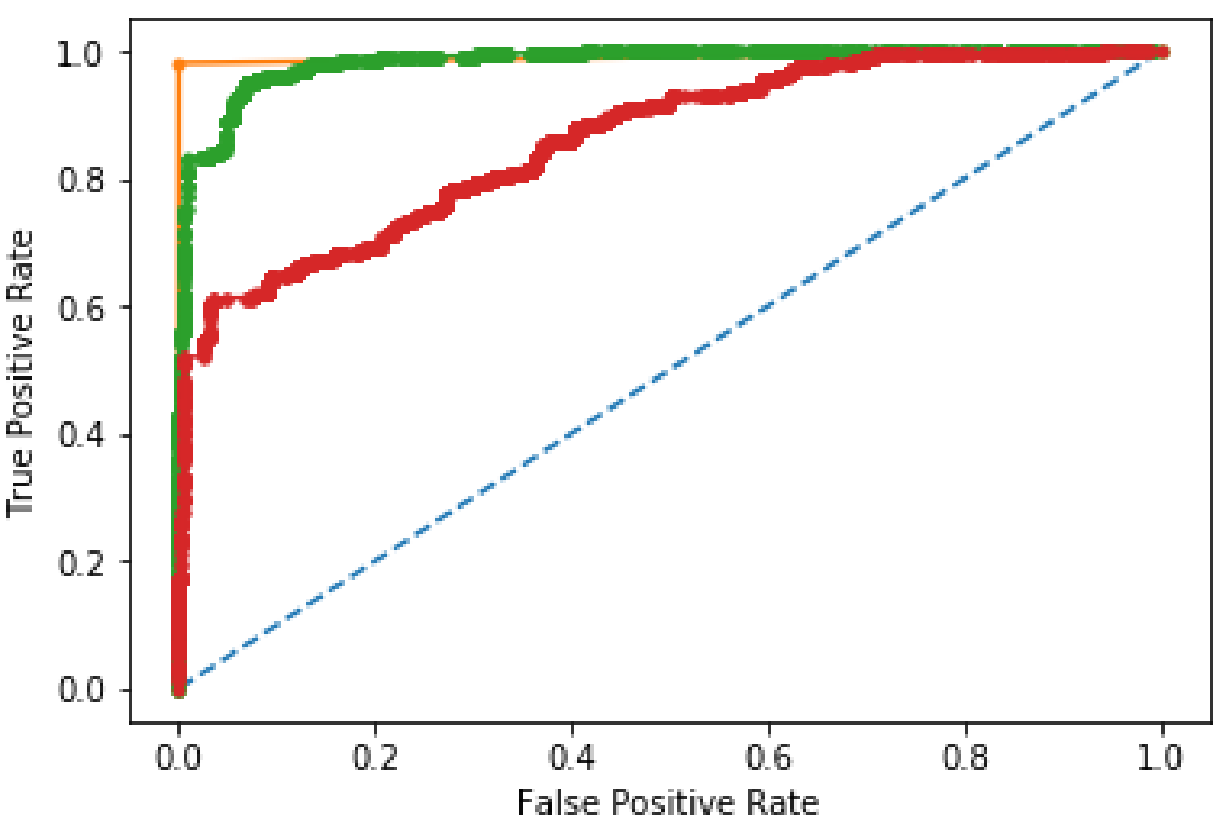

Figure 2: The ROC curves of the three algorithms' classification results

Here, the area under curve (AUC) for Naïve Bayes is 0.86 and its accuracy in detection is 69.5%, whereas for Logistic regression we found the AUC is 0.97 and its accuracy is 93.7%. The AUC for Artificial Neural Network is 0.99 and its accuracy in detecting malware is 98.6%, which is better than the other two algorithms.

## V. Conclusion and Future Works

In the proposed model, multiple methods such as naive Bayes and logistic regression classifiers are used along with two hidden layers, Cross entropy and Adam optimization algorithm. The dataset used in the proposed model had 100000 samples and the dataset was labeled. For logistic regression, the accuracy score is 93.745% for 100000 samples, whereas Naïve Bayes algorithm achieves 69.515%. ANN algorithm is used for malware detection and achieves 99.56% accuracy. Three methods are compared that were used to detect malware. After comparing all three methods, we found that ANN (Artificial Neural Network) got the highest accuracy of 99.56% whereas TPR is 0.99 and FPR is 0.021. With an AUC of 0.99, ANN is better than other algorithms as it is an incredible information-driven, self-adaptable, flexible computational instrument having the capacity of capturing nonlinear and complex fundamental attributes of any physical process with a high level of accuracy. In any case, to completely assess the reasonableness of our methodology, a lot more research needs to be conducted. While our preliminary outcomes are promising, more work is needed to improve the technique and accuracy of our proposed model by using a deep learning method.